\documentclass[aps,prc,amsmath,amssymb]{revtex4}
\usepackage{graphicx,amsmath}

\begin{document}

\title{Designing, Building, Measuring and Testing a Constant Equivalent Fall Height Terrain Park Jump}
\author{Nicola Petrone}
\author{Matteo Cognolato}
\affiliation{Department of Industrial Engineering, University of Padova, Padova, Italy}
\author{James A. McNeil}
\affiliation{ Department of Physics, Colorado School of Mines, Golden, CO, 80401, USA}
\author{Mont Hubbard}
\affiliation{ Department of Mechanical and Aerospace Engineering, University of California, Davis, CA, 95616 USA}

\begin{abstract}Previous work has presented both a theoretical foundation for designing terrain park jumps that control landing impact and computer software to accomplish this task.  US ski resorts have been reluctant to adopt this more engineered approach to jump design, in part due to questions of feasibility. The present study demonstrates this feasibility. It describes the design, construction, measurement and experimental testing of such a jump. It improves on previous efforts with more complete instrumentation, a larger range of jump distances, and a new method  for combining jumper- and board-mounted accelerometer data to estimate equivalent fall height, a measure of impact severity. It unequivocally demonstrates the efficacy of the engineering design approach, namely that it is possible and practical to design and build free style terrain park jumps with landing surface shapes that control for landing impact as predicted by the theory.
 \end{abstract}

\date{September 30, 2016} 
\maketitle

\section{Introduction} 

Snow sports over the past 25 years have seen an evolution toward freestyle skiing and snowboarding which are characterized by aerials occurring in snow parks hosting dedicated jumping features~\cite{NSAA2011}.  Over this period there has also been a documented increase in injuries associated with jumping, some of which involve very large personal and financial ramifications.  Specifically, Yamakawa, et al.~\cite{Yamakawa2001} reported an increased incidence of spine and head injuries in snowboarders compared to skiers, particularly when jumping is involved. Goulet, et al.~\cite{Goulet2005} found that the severity of an injury increased dramatically when the injury was sustained in a terrain park compared to other areas of the resort. They report that the odds ratio for a skier to experience a severe injury increased dramatically from 1.00 (outside terrain park) to 1.50 (within terrain park).   While Moffat, et al.~\cite{Moffat2009} found that upper extremity injuries were twice as likely inside terrain parks as outside, contrary to Goulet et al.~\cite{Goulet2005} , they could find no difference in the severity of injuries (requiring emergency evacuation) sustained within or outside of terrain parks.  (Moffat offered no explanation for this discrepancy.)   However, a later study by Brooks, et al.~\cite{Brooks2010} confirmed the findings of Goulet, et al. by noting that injuries inside snow parks were more severe than those outside the parks with the majority (60\%) involving falls.  They further speculate that the increase in spinal injuries between 2000-2005, particularly among snowboarders, may be due to the advent of aerials in terrain parks.  Other useful references include the early study of spinal cord injuries by Tarazi, et al.~\cite{Tarazi1999} as well as the literature reviews by Ackery et al.~\cite{Ackery2007} in 2007 on spinal cord and traumatic brain injuries, by Hagel~\cite{Hagel2005} and Sutherland, et al.~\cite{Sutherland1996} comparing snowboarding and skiing injuries and other studies with a specific focus on SCIs ~\cite{Meyers1999} -\cite{Jackson2004}.  Collisions with obstacles (trees in particular) present the major hazard of severe injury to riders outside of terrain parks while jumps present the major severe injury hazard to those within terrain parks.  

A carefully designed snow park layout together with the adoption and application of appropriate safety barriers suggested by Petrone ~\cite{Petrone2012} can somewhat mitigate the risk of collisions. Russell~\cite{Russell2011} has offered several other strategies to reduce risk of jump injury,  including wearing protective equipment (helmets, back guards, etc.), teaching falling techniques, promoting progressive training, reducing the size of jumps, and controlling for speed by limiting the in-run. Similarly, engineering design of jumps through the control of landing surface shape can mitigate risks of injury to jumpers
by controlling landing impact directly. The feasibility of fabrication and performance of such jumps are the subject of this paper.

Social costs from severe injuries can be significant.  In addition to the personal tragedy, the lifetime cost to treat a young victim of a catastrophic spinal cord injury (SCI) can be more than \$10M/event~\cite{Dijkers1995} -\cite{Salvini2007}.   The principal contributing factors to SCI on jumps are landing on the head/neck (i.e. inverted) and landing with enough impact energy to have a high probability of spinal damage.  In the litigious U. S. legal environment much effort is expended attempting to assign responsibility for accidents involving SCI to either the rider or the resort. The skiing industry  has largely relied on anticipatory releases relieving them of responsibility for injury, even due to their own intentional negligence.  In a recent case~\cite{Bagley2014}, however, the Oregon Supreme Court declared that the liability waivers normally required when buying a lift ticket are procedurally and substantively ÒunconscionableÓ,  and therefore unenforceable.  The court further ruled~\cite{Bagley2014} that resorts have a ''duty of care'' in the creation of snow park jumps because they have ''the expertise and opportunity - indeed the common law duty- to foresee and avoid unreasonable risks of their own creation''.

Among the proponents of assigning responsibility to the rider, some researchers have emphasized the importance of the biomechanics of jumping, especially the importance of avoiding an inverted position at landing.  As reported by Scher, et al.~\cite{Scher2015a},  drops from even relatively small heights can result in neck loads that can lead to spinal injury if the rider is in a particularly unfortunate head-down position.  There is no question that riders occasionally make mistakes that put them at risk; nevertheless an engineering approach could allow the construction of jumps that reduce the likelihood that a mistake will result in a catastrophic outcome.  For example, since the {\it likelihood} of spinal injury necessarily increases as the energy of the impact increases, limiting the landing impact energy will reduce the risk of injury. When discussing the kinetic energy of impact, it is common to define an ``equivalent fall height'' (EFH), the kinetic energy associated with the landing velocity component perpendicular to the landing surface divided by $ m g$, where $m$ is the mass and $g$ is the acceleration of gravity. Similarly, avoiding curved takeoffs with natural inverting properties will reduce the probability that a rider will be in an inverted position on landing. 

To address these two issues, engineered jump designs have been proposed by Hubbard, McNeil and co-workers that limit the energy dissipated at impact by designing the shape of the landing surface appropriately~\cite{Hubbard2009}-\cite{Levy2015}, and that reduce inversion risk by limiting curvature late near the end of the takeoff ramp~\cite{McNeil2011c}.   

Scher, et al.~\cite{Scher2015b} have continued to question the emphasis on EFH by noting that there is another injury mode, namely, the ``back-edge-catch'', whose severity increases with the tangential component of the landing velocity.  When the overall kinetic energy is fixed, reducing the EFH (i.e. the energy associated with the normal component) necessarily increases the tangential component.   While technically true, the argument is unpersuasive because the energy scales of the two components are so different that no significant reduction in the energy available to a back-edge-catch accident can be accomplished by increasing the vertical impact to any value that can be absorbed without falling.  For example, a recent jump by a professional athlete executing a ``triple cork'' resulted in a total landing energy of about 16 m (expressed in EFH units). Of this, 1.2 m was in the normal component of velocity and 14.8 m in the tangential.  Minetti, et al.~\cite{Minetti2010} have showed that the maximum normal component that a rider's legs can absorb is only 1.5 m. Therefore, by maximizing the EFH to this value, the lowest value of tangential energy is 14.5 m, a reduction of only 2\%, and even that comes at the cost of an unacceptable increase in the likelihood the jumper will lose control upon landing.  The back-edge-catch accident is a serious matter, but increasing the EFH is not an acceptable strategy to deal with it.

Engineering snow park jumps is further complicated by the malleable properties of snow. In their 2008 Freestyle Terrain Park Notebook~\cite{NSAA2008}, the National Ski Areas Association (NSAA) claim that there is too much variability in snow properties and user factors to allow engineering to be practical.   In rebuttal, Hubbard and Swedberg~\cite{Hubbard2011} have argued that the variability is bounded by physical limits which can be accommodated in the designs.  Furthermore, the constant equivalent fall height (EFH) jumps they advocate are largely insensitive to the commonly cited sources of variability.  Currently, snow park jumps at ski resorts in the U.S. are not engineered, but rather sculpted from the snow based largely on the previous experience of the resort staff.  Some resorts have used ballistic calculators to predict jump trajectories, but ski resorts have not yet embraced an engineering approach in the design of their jumps. This is due to history and lack of technical expertise but also to concerns over liability and questions of feasibility. The ASTM International F27 Committee on Snow Skiing is in the process of developing standards for freestyle terrain park jumps and must also address the issue of the feasibility of quantitative designs.  

While the theory of impact-limited jump surfaces is now well established, until the work reported in this paper it was an open question whether such jumps could in practice be constructed such that impacts were controllable as designed. Our first attempt to study the feasibility of fabricating jumps designed to control landing impacts was conducted at the Tognola Ski Resort in San Martino di Castrozza, Italy in March of 2013~\cite{HMPC2015}.  A medium-sized constant EFH jump was designed and constructed. Its impact performance was tested using jumpers instrumented with accelerometers on both the jumper and the snowboard and compared an existing standard tabletop jump. However, the accelerometer on the jumper failed which made the extraction of EFH values from just the single snowboard accelerometer difficult and prone to large uncertainties. The current study corrects this problem and improves on the earlier work by expanding the range of jump distances tested as well.  

This paper is organized as follows. In Section II we review the theory connecting landing surface geometry to EFH, the process used to create the test jump design, and the methods used to construct and measure the shape of the test jump built in San Vito, Italy. In Section III we present the experimental components used to measure the impact performance of the jump using instrumented professional jumpers. In Section IV we present the results of the data analysis, followed by a discussion of results in Section V and conclusions in Section VI. We conclude that jumps that control impact performance can be readily designed and constructed in practice and that the limited impact can be measured to verify the design by using accelerometers mounted to the jumper and the board.

\section{Designing, Building, and Measuring the Constant EFH Jump} 

\subsection{Calculation of Landing Surface Shape}

Following the methods outlined in McNeil, Hubbard, and Swedberg~\cite{MHS2012},  a constant EFH jump was designed.  First, we briefly review the theory connecting landing surface geometry to impact performance.

Large EHF arises from the impulse from the snow required to annul the component of jumper velocity perpendicular to the snow surface at landing. EFH can be made small, in general, by orienting the snow surface to be nearly parallel to the jumper velocity vector at landing. Under conditions where air drag can be neglected, this results in a differential equation for the shape, $y_L(x)$, of a general landing surface as a function of an arbitrary equivalent fall height, $h(x)$, and which is given by (Eq.(20) in Ref.~\cite{MHS2012})
\begin{equation}
y_L'(x)=\tan\Biggl[\tan^{-1}( \frac{2y_L(x)}{x}-\tan\theta_T) + 
\sin^{-1}\sqrt\frac{h(x)}{\frac{x^2}{4(x\tan\theta_T-y_L(x))\cos^2\theta_T}-y_L(x)} \Biggr].
\label{yprime}
\end{equation}
where $y$ and $x$ are the vertical and horizontal distances, respectively,  measured in a coordinate system with origin at the takeoff point. In  Eq.~\ref{yprime}, first shown in Ref.~\cite{Hubbard2009}, the first derivative of the landing surface function $y_L(x)$ is a function of two variables ($x$ and $y_L(x)$) and three parameters ($g$, $\theta_T$ and $h(x)$ ): the acceleration of gravity, the takeoff angle and the EFH.  For the specific case of \textit{constant} EFH, the general EFH function is replaced by the desired constant value, $h(x)\rightarrow h$.  Solutions to Eq.~\ref{yprime} with constant $h$  will be referred to as ``constant EFH" surfaces. 
One key attribute of any surface, $y_L(x)$, that satisfies this differential equation is that a jumper will experience an impact with a value of EFH equal to $h$, \textit{no matter the takeoff speed $v_0$ and consequent landing location}. This surface has had its landing impact property designed into it through the specification of  $h$ that parametrizes the surface and the result is  
insensitive to takeoff velocity, $v_0$.   As noted above, Eq.~\ref{yprime} is general in that the EFH could in fact be a function $h(x)$ of $x$, thereby giving the jump designer freedom to specify the EFH everywhere.

To find specific instances of constant EFH surfaces one needs to solve Eq.~\ref{yprime} numerically. First, one must specify the values of the external parameters $\theta_T$  and $h$, and, since it is a \textit{first order} differential 
equation, one must also choose a specific boundary condition $y_L(x_F)$ at some
value of $x_F$.  For technical reasons~\cite{Hubbard2009,Swedberg2010} related to the behaviour of the equation at small values of $x$, it is 
desirable to integrate Eq.~\ref{yprime} backward, rather than forward, in $x$. We implement this by taking $x_F$ to be the terminal point for the constant EFH surface.  The arbitrariness of the boundary condition means that for fixed $h$ there is an infinite number of such solutions parametrized by $y_L(x_F)$.  

As shown in Ref.~\cite{Levy2015}, the relation in Eq.~\ref{yprime} can also be inverted. By solving for $h(x)$ from Eq.~\ref{yprime}, one obtains an expression for the EFH everywhere on the landing surface as a function of the specified surface shape, $y_L(x)$, and its derivative,
\begin{equation}
h(x)=\left[\frac{x^2}{4(x\tan\theta_T-y_L(x))\cos^2\theta_T}-y_L(x)\right]\sin^2\left[\tan^{-1}\left(\frac{2 y_L(x)}{x}-\tan\theta_T\right) - \tan^{-1}y'_L(x) \right].
\label{general_EFH}
\end{equation} 

The two relations in Eq.~\ref{yprime} and Eq.~\ref{general_EFH} show the inextricable connection between the landing surface shape $y_L(x)$ and the ensuing EFH $h(x)$ that this shape produces. Given one of the functions, the other can be calculated. This is useful not only as employed in this section to design (using Eq.~\ref{yprime}) the surface shape to produce a desired EFH, but also as a way to estimate the EFH (with Eq.~\ref{general_EFH}) given a calculated or measured surface shape as is done in section IV below.

The section of the ski resort hosting the jump had an approximately constant parent slope of 17$^\circ$.  For this proof-of-principle experiment we designed a medium-sized constant EFH jump of about 14 m (horizontal) length, with a takeoff angle of 10$^\circ$ and a constant EFH of 0.5 m. The terminal conditions $y_L(x_F)$ used to integrate the constant EFH differential equation (Eq.~\ref{yprime}) were tuned until the resulting jump surface intersected the parent slope at around 14 m and the angle of the landing surface at that point was less than 30$^\circ$ (assumed to be the practical limit for the snowcat).  The profile of the resulting jump design is shown in red in Fig. 4.

\begin{figure}
\vskip 0 in
\includegraphics[width=.8\textwidth]{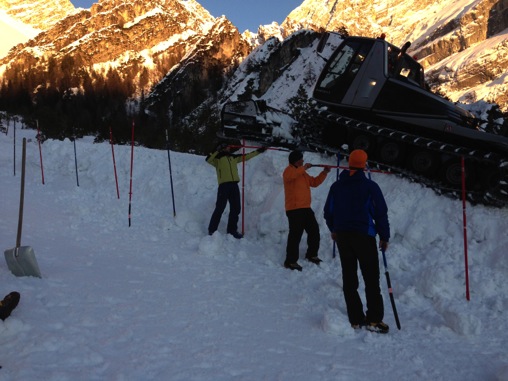}
\vskip 0 in
\caption{In the process of constructing the constant EFH jump on the Antelao pitch of the San Vito ski resort.} 
\label{Figure1} 
\end{figure}
\begin{figure}
\vskip 0 in
\includegraphics[width=.8\textwidth]{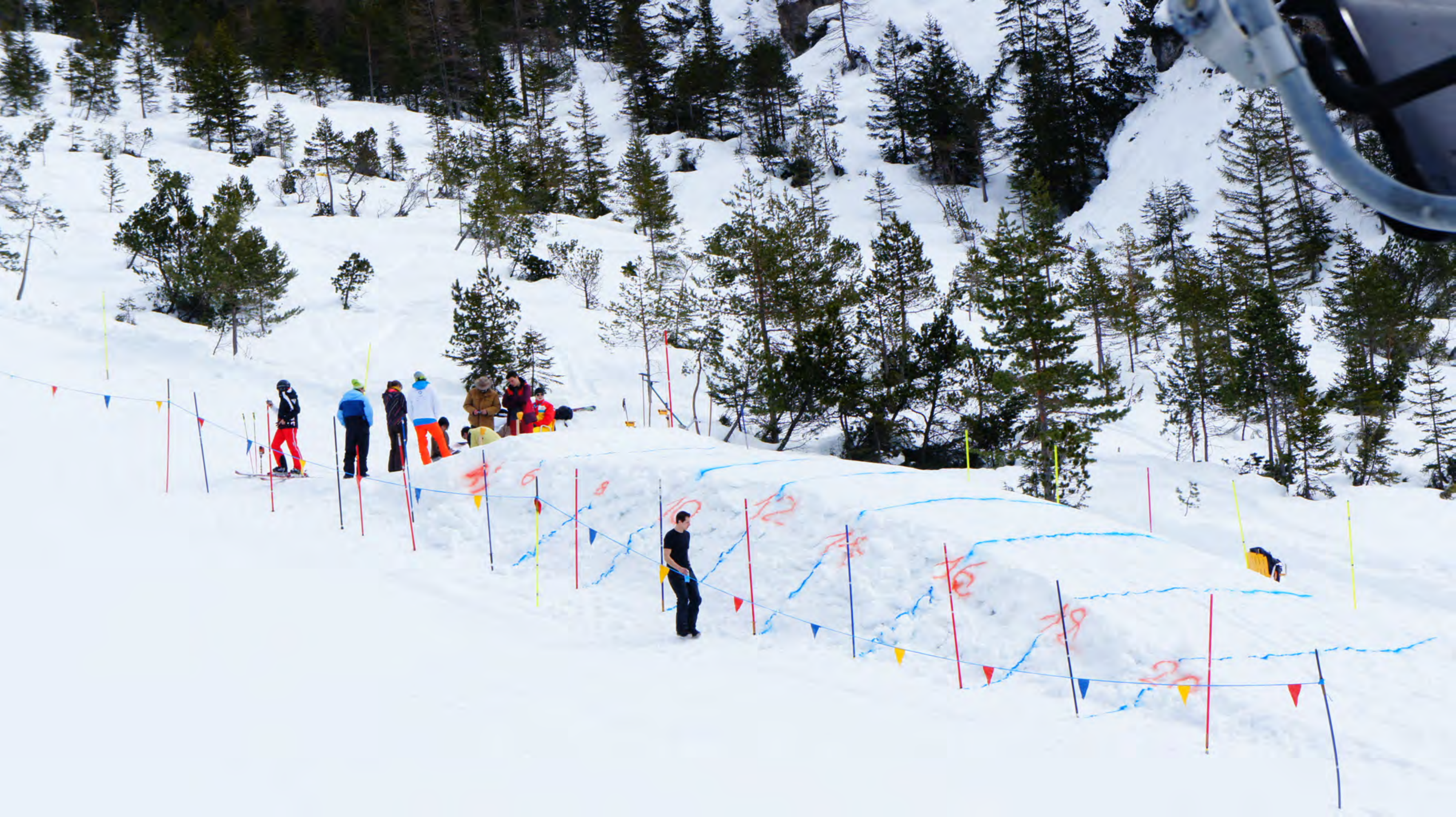}
\vskip 0 in
\caption{Finished view of the constant EFH jump.} 
\label{Figure2} 
\end{figure}

\subsection{Construction}
  
The jump was sited on the Antelao pitch at the San Vito ski resort in San Vito di Cadore, Italy. The San Vito ski resort does not host a snow park so the slope maintenance staff were not well-practiced in building jumps.   
To facilitate the construction in accordance with the design, a set of slalom poles was prepared and placed vertically in the snow as a guide for the snowcat operator, with the jump surface height above the parent slope marked with tape. The poles were placed on the parent slope adjacent to the jump site at 2 m intervals (Fig. 1).  Figure 1 shows the jump under construction.   For safety reasons it was constructed in the late afternoon after the ski resort was closed to patrons. The temperature was slightly above freezing after a sunny day which meant the snow was wet and easily compacted.  The base was sufficiently deep that the jump was able to be constructed using only snow collected from the base. The basic jump landing shape was constructed by the resort staff in about 12 passes using a Prinoth snow groomer.  On the final pass the tiller was used to smooth and groom the landing surface.  The sides of the landing were skirted to allow marking the jump for filming and observational purposes. (The jump was not open to the public.) 

Once the basic landing surface was finished, the takeoff was constructed and measured to have the correct takeoff angle of 10$^\circ$. Some care was taken to insure that the transition curvature did not produce excessive compression, and that the last 2 m of the takeoff were straight to avoid any potential inadvertent inversion hazard. The approach was then groomed and marked with poles at 5 m intervals from the takeoff point to mark the various starting points to be used during the testing.  The entire jump was constructed in about three hours and comprised an approximate volume 100 $m^3$ of snow above the parent surface. The final shape is shown in Fig. 2 with the poles and marking lines on the landing surface.

\subsection{Measurement of the Jump Profile}

The centerline profiles of the jump was measured using special-purpose device that is rolled down the jump centerline, simultaneously measuring distance along the surface using an ATM103 rotary encoder and slope angle using a three axis ADXL345 accelerometer.  Each measuring device was controlled by an Arduino UNO microprocessor board which also stored the data on an SD card for later extraction and processing.  The distance was measured by the rotary encoder by tracking the angle that the wheel of the device had turned. Care was taken to insure the wheel did not slip while the measurement was made. However, we noticed occasional buildup of snow on the wheel which resulted in a systematic underestimation of the distance by about 3\% based on comparison with tape-measured distances on the slope.  The data was corrected to remove this error. The incline angle was measured using two axes of the ADXL345.   To obtain the zero angle offset bias calibration for each pass the device was pointed uphill and then downhill at the same location recording the angle in each case to obtain the zero offset.

The jump profile was obtained from the angle-distance data, $\{\theta,s\}$, using a {\it Mathematica} Notebook. The distance data $s$ were quite stable and reliably reproduced in each sample.  However, the angle data ${\theta}$ derived from the accelerometer were somewhat noisy and included systematic errors resulting in differences between samples. The various sample runs were reconciled with each other by insuring that each run terminated at the same point at the lip of the takeoff.  The angle-distance data was fit to a polynomial in each segment of the jump (the approach to takeoff and the takeoff to end of the landing surface).  Fig. 3 shows an example set of $\{\theta,s\}$ data.  From a polynomial fit to  $\{\theta,s\}$ the x-y profile of the jump landing surface, $y_L(x)$, could be calculated. The result of this process is shown by the blue curve in Fig. 4.  One can see that the measured shape (blue) for the constant EFH jump agrees reasonably well with the design shape (red) with maximum differences of approximately 0.2 m.  

\begin{figure}
\vskip 0 in
\includegraphics[width=.6\textwidth]{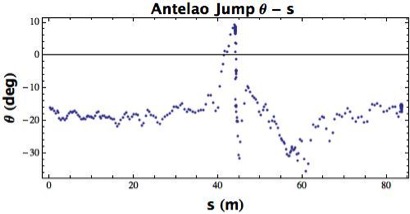}
\vskip 0 in
\caption{Park Profiler $\theta-s$ data for the constant EFH jump.} 
\label{Figure3} 
\end{figure}

\begin{figure}
\vskip 0 in
\includegraphics[width=.8\textwidth]{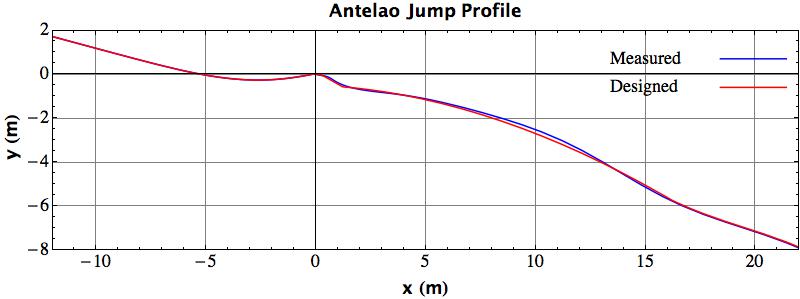}
\hskip 0 in (a) \hskip 3.1in (b)

\caption{Comparison of the designed (red) and measured (blue) jump profiles for the constant EFH jump.} 
\label{Figure4} 
\end{figure}

\section{Impact Behavior of the Jump}

In a normal landing when a jumper impacts the landing surface, the component  of his or her velocity parallel to the surface, $v_{||}$,  can continue relatively unchanged (except for friction and snow deformation). But the velocity component perpendicular to the surface $\Delta v_\perp$ must be cancelled by a force impulse from the snow on the skis or snowboard~\cite{Hubbard2009, MHS2012}, so that after a successful jump the jumper will be traveling parallel to the surface.  The change in the perpendicular component of velocity $\Delta v_\perp$ can be obtained by integrating the perpendicular component of acceleration with respect to time over the duration of the impact. By definition, the \textit{equivalent fall height}  is the  distance $h$ the jumper would need to fall vertically onto a horizontal surface to experience the same impulse. The EFH, $h$, can be calculated from the velocity change $\Delta v_\perp$ through the relation $h=\Delta v_\perp^2/(2g)$, where $g$ is the acceleration of gravity~\cite{Hubbard2009, MHS2012}. Thus, measurement of the EFH for a jump is equivalent to measuring the change in the component of velocity perpendicular to the landing surface. We do this by integrating all components of the acceleration, measured using accelerometers mounted to the board and on the jumper.  The board and jumper must sustain equal changes in velocity to be traveling at the same speed after impact.

The jumper and board were instrumented with two three-axis accelerometers:  first a triaxial HBM SOMAT SAPE-HLS-3010-2 (full scale $\pm$ 500g, 10 mV/g sensitivity, 10 kHz 3 dB bandpass) was placed on the jumper's lower back, between the two PSIS, just over the sacrum bone, corresponding to the vertical standing location of the center of mass (COM) (Fig. 5a). A second triaxial accelerometer was placed  between the jumper's feet, near the center of the snowboard, as shown in Fig. 5b. 
The jumper's accelerometer had the z-axis along the spine while the snowboard had the z-axis perpendicular to the board.
Data were acquired at 5 kHz by the HBM SOMAT e-DAQ lite datalogger (24-bit resolution, synchronous sampling on each channel) placed in a backpack worn by the jumper. The total additional mass was about 3 kg.  The take off velocity was recorded by two MICROGATE POLIFEMO photo-gates spanning 1 m distance, placed at the last portion of the takeoff ramp.  To provide wide range of takeoff speeds and landing distances, different starting points were marked: from 10 m to 40 m at 5 m intervals up the run-in of the jump.  The landing surface was marked in order to determine visually the length of each jump: the marking lines were made at 2 m intervals over the jump surface and closely observed by two observers in order to detect the point of first contact on the snow.   Furthermore, each jump was recorded by a 50 fps video camera, set up approximately 30 m away from the jump providing a good profile view. Starting distance,  takeoff speed and landing location were recorded for each jump.

To simplify the analysis and have the experiment comport with the assumption of the theory that the takeoff velocity vector is parallel to the takeoff ramp,  jumpers were instructed not to ``pop'' (jump) or rotate at takeoff.  Starting points were varied from 10 m to 40 m up the run-in to provide a wide range of takeoff speeds and landing distances. A total of 21 instrumented jumps were made; 6 on Day 1 of testing and 15 on Day 2.  After Day 1 the jumper's y-axis accelerometer was found not to be working but this was repaired overnight so only the data from Day 2 were analyzed.

\begin{figure}
\vskip 0 in
\includegraphics[width=.4\textwidth]{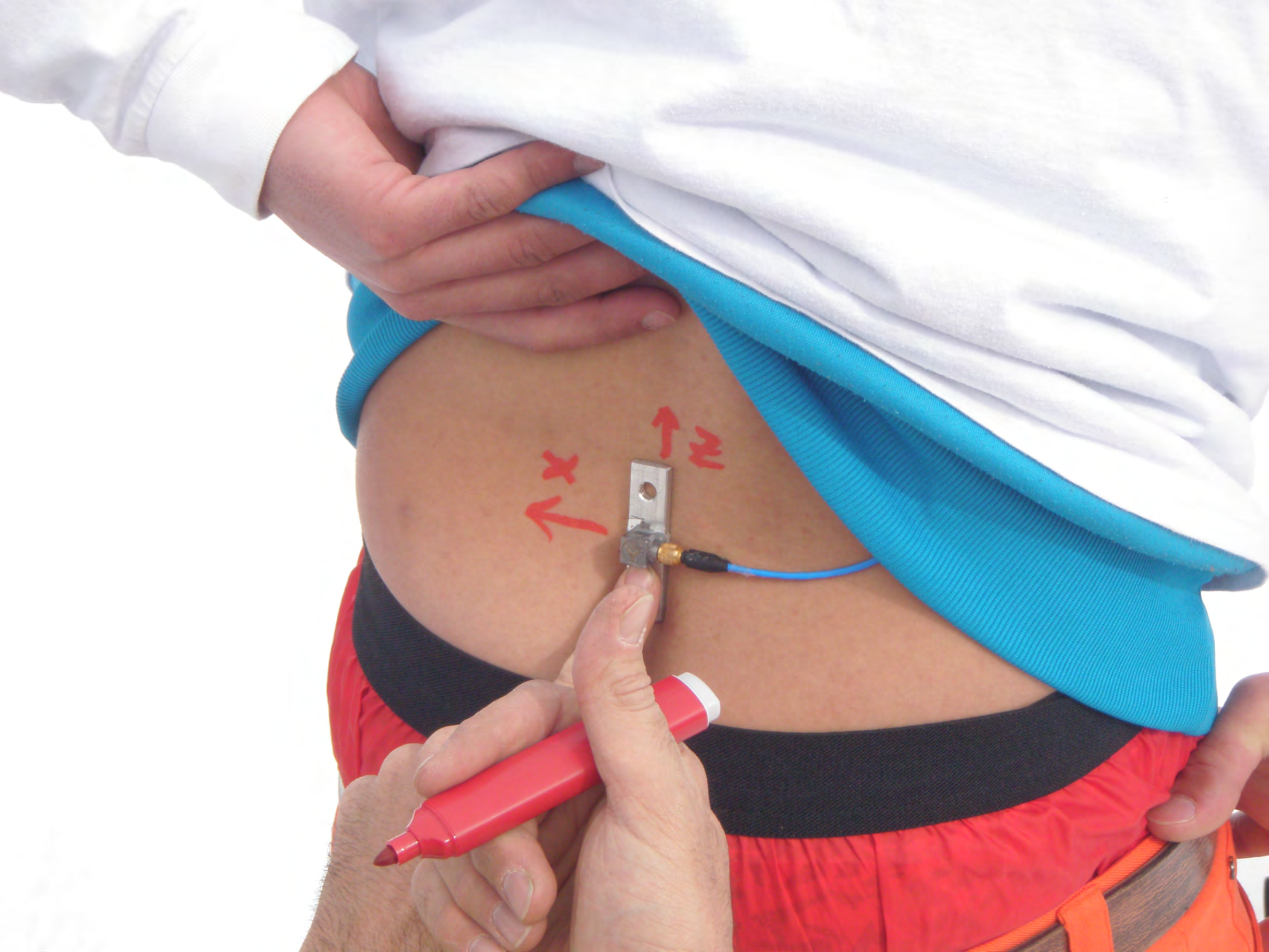}
\hskip 0.4 in
\includegraphics[width=.4\textwidth]{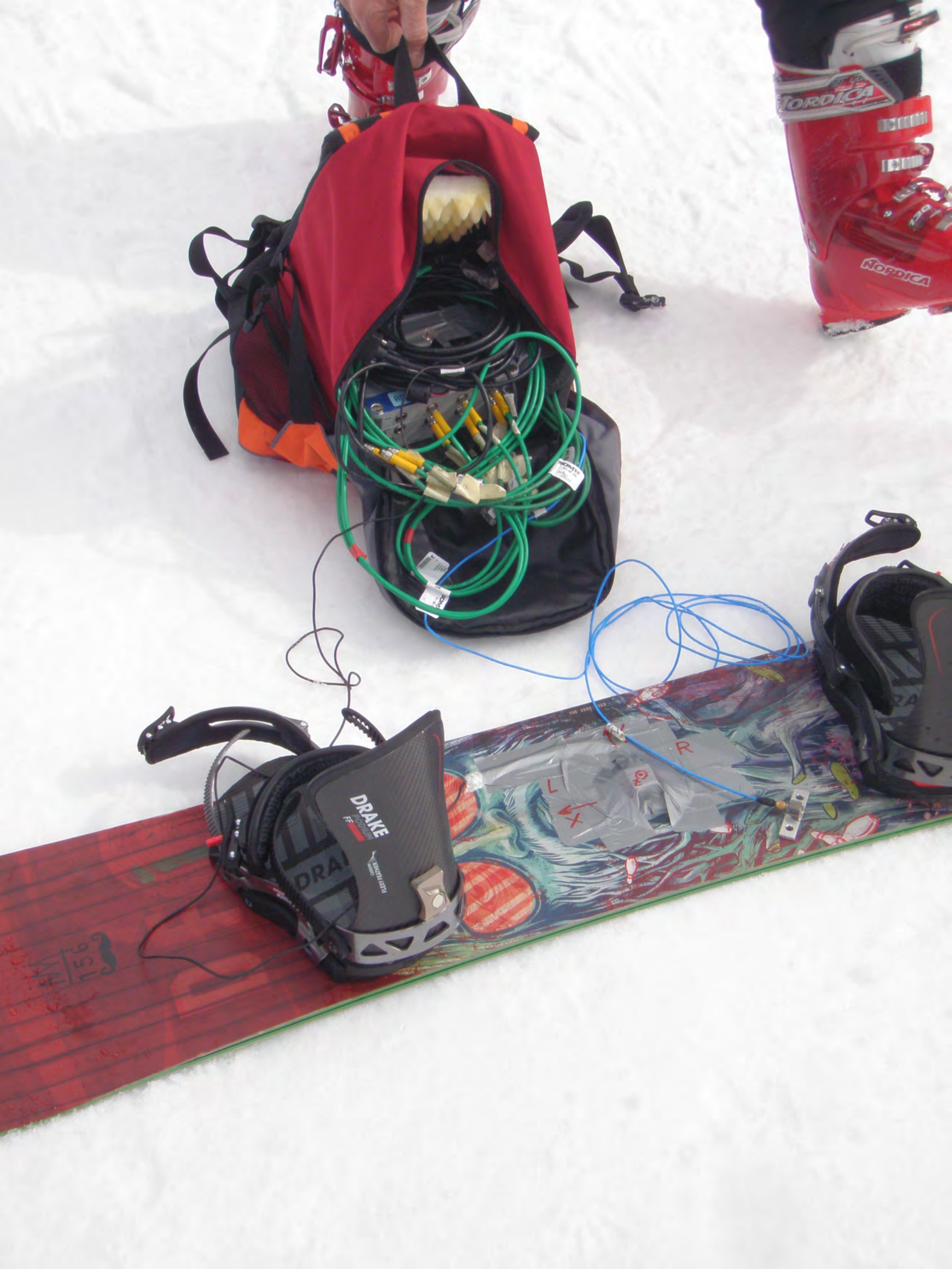}
\vskip 0 in
\hskip 0 in (a) \hskip 3.1in (b)
\caption{Location of accelerometer on the jumper (a) and the snow board (b). } 
\label{Figure5} 
\end{figure}


\begin{figure}
\vskip 0 in
\includegraphics[width=.4\textwidth]{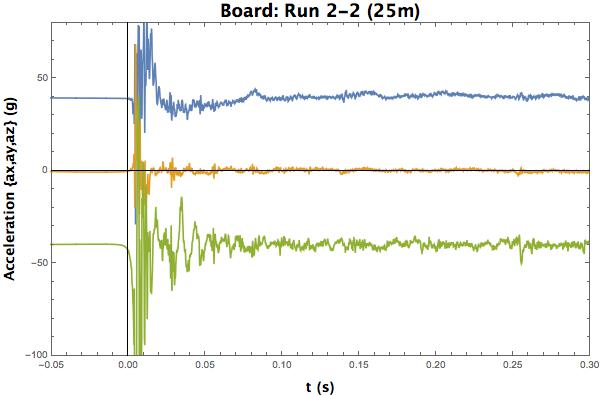}
\hskip .4 in
\includegraphics[width=.4\textwidth]{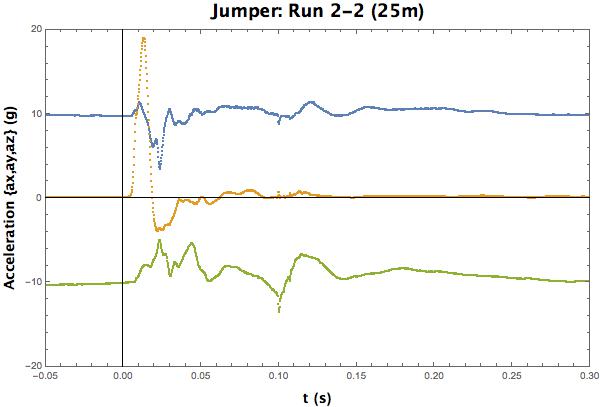}
\vskip 0 in
\hskip 0 in (a) \hskip 3.1in (b)
\caption{Example  accelerometer data (a) for the board ($a_x$ upper line (blue), $a_y$  middle line (yellow),  $a_z$ lower line (green)) and   (b) for the jumper ($a_x$ upper line (blue), $a_y$  middle line (yellow),  $a_z$ lower line (green)). In both figures the quiet flight period is clearly visible just before impact (the vertical straight line). To improve readability, the three $xyz$ traces are separated in the two figures by 50g and 10g, respectively. Note that the board experiences much larger accelerations for a shorter time period with more high frequency content than those of the jumper COM. } 
\label{Figure6} 
\end{figure}

\begin{figure}
\vskip 0 in
\includegraphics[width=.6\textwidth]{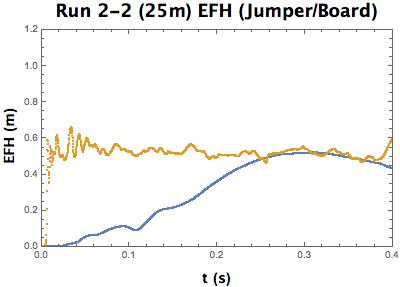}
\vskip 0 in
\caption{Plot of EFH for one jump obtained by integrating the resultant board (upper curve (yellow)) and jumper (lower curve (blue)) accelerations with time after impact. The effective EFH for the jump was estimated as the value where the two curves first intersect, in this case about 0.5 m at 0.25 s. } 
\label{Figure7} 
\end{figure}

\section{Results}

Figure 6 shows a typical example of the three components of the jumper and snowboard acceleration versus time just after impact. To improve readability, the three $xyz$ traces are separated in Figs. 6a and 6b by 50g and 10g, respectively. One can clearly see the quiet region of flight just prior to impact.  By zooming in on the region near the impact we identified the board's instant of impact and shifted the time to make that the origin. This time shift is arbitrary and is used only to identify the start time for integrating the acceleration to obtain the change in velocity.  As expected, the jumper's accelerations are much smaller in magnitude but of longer duration than those of the snowboard.  

Figure 7 shows the change in the snowboard and jumper velocities, $\Delta v=\sqrt{\Delta v_x^2+\Delta v_y^2+\Delta v_z^2}$ obtained by integrating the corresponding resultant accelerometer signals after impact as functions of time. We express these in ``EFH'' units, namely $\Delta v^2/(2 g)$.  The change in velocity of the board is dominated by the perpendicular component $\Delta v_z$, while for the jumper the y-component is largest, but all components appear to contribute.  This is what one would expect from a snowboarder landing in a normal (sideways to the downhill direction) and bending forward to absorb the impact.  For a typical (flat board) landing one sees a steep rise in $\Delta v$ for the snowboard which then drops briefly below the axis before returning to an approximately constant value. This appears to represent a small bounce or recoil upon impact lasting approximately 0.05 s.   We consider the impact concluded when the snowboard and jumper velocity changes are equal, that is when they are moving again at the same speed.  In the example shown in Figure 7, this occurs approximately 0.25 s after impact, but this time varied slightly from jump to jump, depending on how the jumper used the legs to cushion the impact.  To obtain a value of EFH for each jump, we average the board and jumper $\Delta v$ over a small time interval (0.02 s) around the time the two become equal. We calculate the value of EFH using $h=\Delta v^2/(2g)$.  As an estimate of the error, we average the absolute value of the difference over the same time interval.    

Since we integrate all components of the acceleration, the change in the component of velocity parallel to the surface is included as well.  To assign the result to the actual EFH assumes that the component of velocity parallel to the surface does not change significantly during the landing impact leading to a systematic over-estimation of EFH which is defined in terms of the change in the normal component of velocity only.  This effect induces a systematic offset of approximately $2\mu \simeq$ 12\%, using an estimated friction coefficient of $\mu\simeq 0.06$~\cite{McNeil2011b}.  

Table 1 summarizes the data for the 15 jumps on Day 2 showing the run number, distance jumped, the experimental value of EFH, and the deviation from the EFH expected from the jump profile and impact location for each jump.  The average magnitude of the deviation of the experimental from the theoretically predicted EFH values is 0.17 m. The error in the jump distance was estimated to be about 0.2 m based on variations in the visual observations of the jump distance made by the two observers. 

\begin{table*}
\caption{Summary of jump data}
\label{Tab:1}  
\begin{tabular}{ccccc}
\hline\noalign{\smallskip}
Run~~~~~ & Distance (m) & ~~EFH(exp) (m) & $~~|\pm\Delta$EFH(exp)$|$  (m)  \\
\noalign{\smallskip}\hline\noalign{\smallskip}
2-1 &  6.3  &  0.49 & 0.10  \\
2-2 &  6.0  &  0.50 & 0.07 \\
2-3 &  5.9  & 0.50 &  0.10  \\
2-4 &  2.9  & 0.51 & 0.27  \\
2-5 &  2.9 &  0.65 & 0.32  \\
2-6 &  3.0 & 0.48 & 0.08  \\
2-7 &   1.4 & 0.42 & 0.20 \\
2-8 &   1.5 & 0.36 & 0.15 \\
2-9 &   1.5 & 0.48 & 0.23 \\
2-10 &  9.4 & 0.40 & 0.08 \\
2-11 &  9.3 & 0.33 & 0.06  \\
2-12 & 10.1 & 0.40 & 0.12 \\
2-13 &   11.5 & 0.40 & 0.20 \\
2-14 & 11.6 & 0.51 & 0.37  \\
2-15 &  12.6 & 0.59 & 0.17\\


\noalign{\smallskip}\hline
\end{tabular}
\end{table*}


\begin{figure}
\vskip 0 in
\includegraphics[width=.6\textwidth]{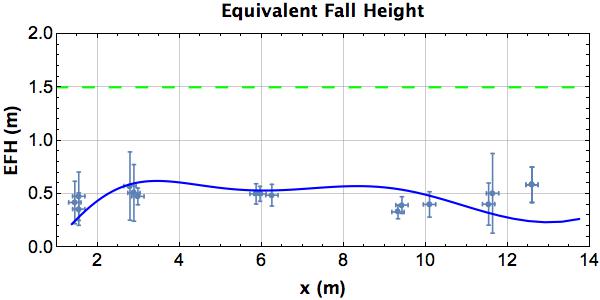}

\caption{Predicted EFH for the measured jump profile (solid-blue) compared with EFH values extracted from the accelerometer data as described in the text and Table I.  Also shown is the USTPC~\cite{USTPC}  limit (dashed-green) for EFH based on experiments with elite ski jumpers by Minetti, et al.~\cite{Minetti2010}.  Error bars indicate best estimates of uncertainties associated with visual estimation of landing distance, $x$, and of uncertainties in the calculation of EFH, $h$.} 
\label{Figure8} 
\end{figure}

The data points in Fig. 8 show the 15 experimental values of EFH extracted from the board and jumper accelerometer data, EFH(exp) from Table I, compared with the EFH function, EFH(theory) (solid-blue line), predicted using Eq.~\ref{general_EFH} and the measured constant EFH landing surface shape, both plotted as a function of horizontal distance jumped.

\section{Discussion} 

Freestyle terrain park jumping offers excitement of aerial maneuvers but at the cost of greater risk of injury.  Of particular concern is the risk of spinal cord injury due to landing with too severe an impact in an inverted position.  Russell~\cite{Russell2011} has suggested several steps to mitigate the risk to jumpers including wearing protective equipment, training jumpers better, reducing the size of jumps, and limiting takeoff speeds.  A complementary approach proposed in Refs.~\cite{MHS2012} and \cite{Levy2015} is to design the jump itself to avoid inadvertent inversions by having straight takeoff ramps and to limit landing impact through control of the shape of the landing surface.  One such solution has been tested in this study through the design and construction of a constant equivalent fall height jump. Such non-traditional jumps are characterized by continuously steepening landing surfaces that may appear to present a challenge to jump builders, especially if the maximum slope exceeds the capabilities of snow groomers. This purpose of this work has been to demonstrate the feasibility of constructing such jumps and to show that they perform as expected in limiting impact.

The impulse sustained by the jumper on landing occurs as the component of impact velocity perpendicular to the landing surface is annulled. This can be measured with accelerometers on the jumper since all objects (the jumper herself, the snowboard, even the poles, etc.) sustain essentially the same net velocity change, the same mass-specific impulse, and the same EFH (that this is true is made particularly clear in the close correspondence of the board and COM curves of Fig. 7 between 0.25 and 0.38 s). To integrate the accelerations to obtain the velocity change and the consequent EFH, however, one needs a time period of integration. In a previous study of the impact performance of standard tabletop and constant equivalent fall height jumps~\cite{HMPC2015}, we had chosen this time somewhat arbitrarily to be 0.2 s. Although this previous crude method provided reasonably accurate results, we believe the more precise method explained in this paper appears to be superior because it relies on the natural time period required to make the board and jumper COM velocity changes equal. 

Figures 6 and 7 portray the accelerations and changes in velocity (expressed as consequent EFH), respectively, as functions of time as they are experienced during a single jump. In both Figs. 6(a) and 6(b),  the quiet flight period is clearly visible just before impact (the vertical straight line). To improve readability, the three $xyz$ traces have been separated in the two figures by 50g and 10g, respectively. Note that the board experiences much larger accelerations for a shorter time with more high frequency content than those of the jumper COM. But Fig. 7 shows clearly that, even though it takes longer (about 0.25 s) for the jumper COM velocity vector to become tangent to the landing surface as a result of the landing being cushioned by the leg muscles (Minetti, et al.~\cite{Minetti2010}), eventually the EFH experienced by the board and COM are the same.

Figure 8 shows how the EFH  experienced by the jumper varies with landing position along the surface. Because of our inability to fabricate the designed landing surface shape as exactly as we had calculated it, even though the desired EFH was 0.5 m, the predicted EFH varied slightly due to these fabrication imperfections. So in some sense it is appropriate to compare the experimentally measured EFH both to the desired value and to the value expected from what was actually built.  While the EFH calculated in a given jump with a given landing position characterizes the impulse received by the jumper during that specific landing, the entire set of such results over a comprehensive range of landing positions characterizes the overall safety of the jump itself. 

An EFH of $h=0.5 m$ corresponds to a relatively gentle landing surface. Values of EFH between 3-10 m have been found after measurement and analysis of landing surfaces in which severe SCI injuries have resulted~\cite{Salvini2007}. So the fact that the measured values of EFH in Fig. 8 differ by as much as 0.2 m from the desired constant value, and even from the expected slightly variable value, is of little concern. Figure 8 shows two things. First we were able to fabricate a landing surface shape that produced an EFH function $h(x)$ close enough to the desired value. Secondly, the actual EFHs experienced by the test jumpers were near those predicted by the theory. 

\section{Summary and Conclusions} 

To test the feasibility of constructing freestyle terrain park jumps that control for landing impact, we designed and built a constant EFH jump at the San Vito ski resort.  We ensured the constructed jump complied with the one designed  by using marked slalom poles to guide the snowcat operator in the construction process. This expediency worked well for this medium-sized jump, but for larger jumps whose height above the parent slope may exceed 2 m, an alternative method will be required.  Cost permitting, the use of ground-based differential GPS would seem to promise the best option. 

Upon completion, the constructed jump profile shape was measured using the ``Park Profiler'',  a device that simultaneously measures the surface inclination and distance.  The fabricated jump shape was measured to be quite close to the designed shape with a maximum excursion of approximately $\Delta y = 0.2$ m throughout the entire range of the jump $0<x<14$ m.   

To test the impact performance of the jump, the actual energy dissipated at impact was measured using jumpers instrumented with three-axis accelerometers on both the snowboard and the jumper. The change in velocity following impact was obtained by integrating the accelerations with respect to time over the duration of the impact, determined by the time when the jumper COM and board resultant velocities experienced the same change. Typical impact times were in the range 0.20 - 0.30 s which agrees with our previous experience at San Martino~\cite{HMPC2015}. The predicted EFH behavior was calculated and found to be close to the design value $h=0.5$ m throughout most of this range with average magnitudes of deviations of about $0.17$ m.  The accelerometer-determined EFH and the theoretical EFH expected from the measured jump profile agreed quite well over the entire range of distances jumped.

Although we do not believe it needs repeating, the experiment described in this paper could be improved by reducing the weight of the data pack which influenced the rider's posture, possibly employing body suits to capture posture during the jump, and by using a more accurate method of determining the landing location. The construction of the designed jump could be improved through the use of differential GPS to guide the snow grooming machine. Nevertheless, this experiment presents compelling evidence that designed jump surfaces that embody low values of EFH  are practical to build and, once built, perform as predicted in limiting landing impact. The jump constructed and measured in this work clearly demonstrates that impact on landing can be controlled through design of the shape of the landing surface.

%


\vfill\eject
\end{document}